# BAYESIAN MODEL OF INDIVIDUAL LEARNING TO CONTROL A MOTOR IMAGERY BCI

C. Annicchiarico[1,2,3], F. Lotte[2], J. Mattout[1,3]

[1] Lyon Neuroscience Research Center, CRNL, INSERM U1028, CNRS UMR5292, Computation, Cognition and Neurophysiology Team, Lyon, France
[2] Inria Center at the University of Bordeaux / LaBRI, France
[3] Université Claude Bernard Lyon 1, Lyon, France

E-mail: come.annicchiarico@inserm.fr

ABSTRACT: The cognitive mechanisms underlying subjects' self-regulation in Brain-Computer Interface (BCI) and neurofeedback (NF) training remain poorly understood. Yet, a mechanistic computational model of each individual learning trajectory is required to improve the reliability of BCI applications. The few existing attempts mostly rely on model-free (reinforcement learning) approaches. Hence, they cannot capture the strategy developed by each subject and neither finely predict their learning curve. In this study, we propose an alternative, model-based approach rooted in cognitive skill learning within the Active Inference framework. We show how BCI training may be framed as an inference problem under high uncertainties. We illustrate the proposed approach on a previously published synthetic Motor Imagery ERD laterality training. We show how simple changes in model parameters allow us to qualitatively match experimental results and account for various subject. In the near future, this approach may provide a powerful computational to model individual skill learning and thus optimize and finely characterize BCI training.

INTRODUCTION

Motor Imagery is one of the most employed non-invasive BCI paradigm due to its potential in stroke rehabilitation and motor control. Event-related desynchronization (ERD) in motor cortices is associated with motor task execution, observation or mental imagery. It is a key biomarker to pick up to interface the brain with an assistive (e.g. neuroprosthetics) or a rehabilitation (e.g. neurofeedback) device. Studies focusing on MI training have demonstrated notable positive outcomes, including enhanced hand dexterity [1] and post-stroke improvements [2], [3]. These interventions capitalize on the overlapping neural pathways between mental imagery and motor execution. Particularly in the context of hemispheric ischemic stroke, some studies have attempted to address motor control deficits [4] by using neurofeedback training to strengthen MI laterality, with some success [5].

Despite those results, the core neuropsychological mechanisms behind subject self-regulation are still poorly understood. Some theoretical approaches [6], [7] have proposed unifying frameworks to describe such processes during BCI or NF training. Among those processes, the nature of subject learning has been the main focus of academic debate [8]. Two views mostly prevail and are in relative opposition. Proponents of operant conditioning reflect a model-free (reinforcement learning) view on how subjects learn during BCI training [9]. A different view that also assume that subjects learn from trial and error, supporters of cognitive skill learning [10], [11], [12] suggest that subject actively build an interaction model of the BCI system in order to reliably interact with it. According to this second view, users learn a skill ("interacting with the BCI") in order to control the interface despite the high levels of uncertainty of the paradigm. This form of learning, more akin to "model-based" reinforcement learning (RL), provides more satisfactory explanations for phenomena such as transfer learning and the effect of metacognition on regulation [7]. The true nature of subject experience during BCI training probably stands between these two views on adaptation, with initial interactions generally driven by RL and progressively building a more complete model-based representation of the system.

The general lack of understanding of the self-regulation mechanisms at play during successful and failed training procedures has prompted the scientific community towards the development of models of subject learning in order to explain and hopefully predict the outcome of BCI training given a particular subject, experimental design, etc. These models have built on the above-described R.L. perspective to leverage difficult credit assignment problems as when learning individual neuron activations under high uncertainty [13], [14]. We argue that in order to model the cognitive dynamics of training and account for its metacognitive and transfer learning dimensions, an explicit modelling of the subject's representations is needed. To our knowledge, such an approach to BCI has barely been tackled. In this work, we show how the Active Inference framework [15] may be leveraged to provide an adequate theoretical and computational ground for developing this modelling strategy. To illustrate our modelling approach, we consider a rich and original study that has implemented a multimodal right-hand Motor Imagery neurofeedback training [16], [17].

## MATERIALS AND METHODS

*Active Inference and BCI training:* Active Inference is a process theory that provides a description of agent perception, action and representational learning as a single joint process based on minimization of (variational) Free Energy [18], [19]. Active Inference is closely related to the predictive coding account of brain function, which posits that the brain entertains and constantly updates a *generative model* $m$ of the environment in order to formulate accurate predictions about its dynamics and guide its actions. To minimize the prediction errors, agents continuously maintain beliefs about the hidden states of their environment and update them with regard to new observations (perceptual inference). This Bayesian process can be further formalized as follows: assuming a set of beliefs $s$ about hidden (causal) states $\hat{s}$ of the environment and given new observations $o$, updated beliefs about those states write:

$$p(s|o,m) = \frac{p(o|s,m)p(s|m)}{p(o|m)} \quad \text{Eq. 1}$$

Variational (Bayesian) inference provides both the model evidence or marginal likelihood $p(o|m)$ and the posterior distribution $p(s|o,m)$. Precisely, the former is maximized and amounts to minimize an approximate energy function, which is a lower bound to the model evidence (ELBO). Importantly, Active Inference makes use of two additional mathematical constructs to implement full representational learning and action. First, it frames this belief updating process as a Hidden Markov process or model (HMM) thus accounting for the temporal evolution of subject's beliefs. Second, it includes action as part of the energy minimization process, turning this HMM graph into a POMDP (Partially Observable Markov Decision Process) (see Figure 1). In other words, free energy, surprise or prediction is not only minimized through belief updating but also by acting upon the environment to make it fit with predictions.

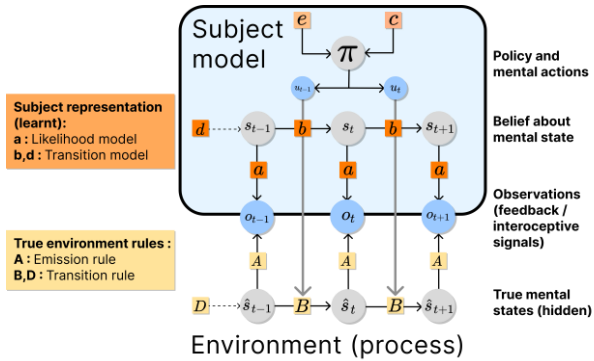

Figure 1: Active Inference (POMDP) canonical model of subject's representation and interaction with a changing environment. This model makes explicit that the agent must infer and navigate the hidden, noisy and partially observable environment based on noisy and sparse information. Importantly, BCI training can be nicely framed with such an POMDP.

Table 1 provides the description of the above model component and parameters in the context of BCI training. Given this formulation, the subject's Free Energy can be minimized in three ways: through perception (inference on hidden states), action (transition between hidden states) and learning (updating model parameters). Under this premise, agent s may pick actions in order to reduce their (expected) free energy on the basis of anticipated future observations, in a way that optimize a trade-off between information seeking (exploration) and reward maximization (exploitation). In this paper, agents plan their future actions by comparing all the plausible action trajectories within a specific temporal horizon [20]. Finally, learning occurs at a slower pace at which subjects update their model parameters. In the discrete state space leveraged by Active Inference, these model parameters are categorical distributions equipped with conjugate Dirichlet priors. Learning occurs through *counting* co-occurrences between state posteriors and observations (likelihood **a**), or transitions between states following a given action (transition **b**), akin an evidence accumulation process [15]. In essence, Active Inference describes the evolution of subject's beliefs $(x, \pi)$ and representations (**a**, **b**, **c**, **d**, **e**) depending on environmental parameters (**A**, **B**, **D**). This translates directly to BCI training where we may cast the feedback provided to the subject as the observations, and the mental states targeted by the training procedure (attention, hand motor imagery level, …) as the true hidden states. The subject tries to reach high levels of positive feedback by learning an accurate representation of the BCI system (**a**, **b**, **d**).

Table 1: Correspondence between Active Inference graph parameters and BCI training elements

| Active Inference parameter | BCI training element |
|---|---|
| **A** | Emission rule: relation between feedback and subject's true mental state |
| **B**, **D** | Transition rule: effect of mental action onto mental states |
| **a** | Subject's belief about the feedback (affected by instructions, experience…) |
| **b**, **d** | Subject's belief about its mental strategies and the effect of its mental actions (idem) |
| **c**, **e** | Subject preferences (towards positive feedback) and habits |
| $\hat{s}$, **s** | True and belief about mental states, respectively |
| **o** | Observations (feedback) |
| **π**, **u** | Subject's mental policy and possible actions |

*A Motor Imagery Neurofeedback training task:* To illustrate our modeling approach, we consider a simplified version of the task implemented by Perronet, Lioi et al. [16], [17]. In their first experiment, (N=10) subjects were instructed to perform kinesthetic right hand motor imagery and to "find their own strategy" in order to control a feedback gauge across 3 x 10 blocks. Each block comprised a 20s rest and a 20s task block. The task was multimodal as both fMRI and EEG data were recorded and the feedback was based on either EEG alone, fMRI alone or both signals (two feedback gauges simultaneously). Importantly, the gauge levels were always based on a measure of lateral asymmetry between left and right motor cortex activities (For EEG: an asymmetry index computed on the normalized difference in $\mu$ (8-12 Hz) band power between C3 and C4, updated every 250 ms; for fMRI: a laterality index as described in [21] , updated every 2 s).

This study is quite unusual, namely because of the two neuroimaging modalities employed. However, it offers an appealing example to model, for at least two reasons. First, the well-defined laterality biomarker permits fairly simple assumptions regarding the subject's self-regulatory process. Second, data availability [17] allows for broad model calibration. In what follows, we propose a computational model of this protocol and provide general predictions regarding long-term training outcomes.

*Modeling Motor Imagery laterality training:* The Motor Imagery neurofeedback loop is formalized as a high uncertainty self-regulation task. The agent trains over $N_{trials}$, each trial being composed of an arbitrary 40 rest and 40 MI timesteps (each timestep corresponding to 2 EEG feedback update for the experimental task). During MI, the agent is given a feedback based on its hidden states and attempts to reach highly rewarding outcomes. No feedback is provided during rest. During the whole training, the agent activity was defined by two hidden states based on electrophysiology : the left and right ERD levels.

$$\begin{cases} \widehat{ERD}_L(t) = \hat{\iota}(t) \cos(\hat{\alpha}(t)) + \epsilon \\ \widehat{ERD}_R(t) = \hat{\iota}(t) \sin(\hat{\alpha}(t)) + \epsilon \end{cases} \quad \text{Eq.2}$$

Where the radius $\hat{\iota}(t) \in [0; 1]$ captures the global ERD strength and the angle $\hat{\alpha}(t) \in [0; \frac{\pi}{2}]$ its lateralization or orientation. $\epsilon$ is a baseline level accounting for spontaneous desynchronizations outside of MI, which we set to 0.01 (weak baseline level).

Agents have no direct observation of these two physiological states that reflect cortical motor excitability and could be associated with mental states such as motor preparation and sensorimotor expectation. In this framework, agents entertain a belief or prior over these states $(i; \alpha)$ and use the feedback provided (see *Emissions*) to update this belief.

We further adopt a discretized, POMDP compatible formulation of our model, considering that $\hat{\iota}, \hat{\alpha}$ (process states) and $i, \alpha$ (model states) can each span a finite set of $Ns$ possible states. For the sake of simplicity, the simulations were conducted with $Ns(\hat{\iota}) = Ns(i) = 4$ {0: null, 1: low, 2: medium, 3: high ERD strength} and $Ns(\hat{\alpha}) = Ns(\alpha) = 5$ {L: left (0) , CL: center-left ($\frac{\pi}{8}$), C: center ($\frac{\pi}{4}$), CR: center-right ($\frac{3\pi}{8}$),  R: right ($\frac{\pi}{2}$) ERD orientation}. Note that discrepancies between the model and the process in terms of state space dimensions could be accounted for and their effect on training simulated within this framework.

*Emissions:* Agents receive outcomes $o_t$ based on their true state $\hat{s}_t = (\hat{\iota}(t), \hat{\alpha}(t))$. This feedback modality (denoted as *AsI*) is based on the laterality of the ERD. It is computed using an asymmetry index between the left and right ERDs, for $\hat{\iota}(t) > 0$ : (Eq.3)

$$\hat{o}_t = AsI(\hat{\alpha}) = \frac{\widehat{ERD}_L(t) - \widehat{ERD}_R(t)}{\widehat{ERD}_L(t) + \widehat{ERD}_R(t)} \in [-1,1]$$

To account for the noise in the biomarker and feature extraction process, the categorical emission matrix **A** encodes the emission rule of the BCI pipeline as a discretized gaussian distribution $Cat(N(\hat{o}_t; \sigma_{proc}))$ with $N_{AsI} = 5$ possible feedback values.

During the experimental task [16], the strength of left ERD was continuously monitored but was not provided as a feedback signal. We mimic this observation channel with a second emission modality (referred to as *L-ERD*) based on the simulated left ERD level. Similarly, these outcomes are not observed by the synthetic subject during training. They are used to compare physiological measurements to model predictions and broadly estimate which parameter values best matched the study results (see *Results*). Just like with the AsI modality, the L-ERD observations are noisy (same noise parameter $\sigma_{proc}$) and discretized so as to take one out of 5 possible values.

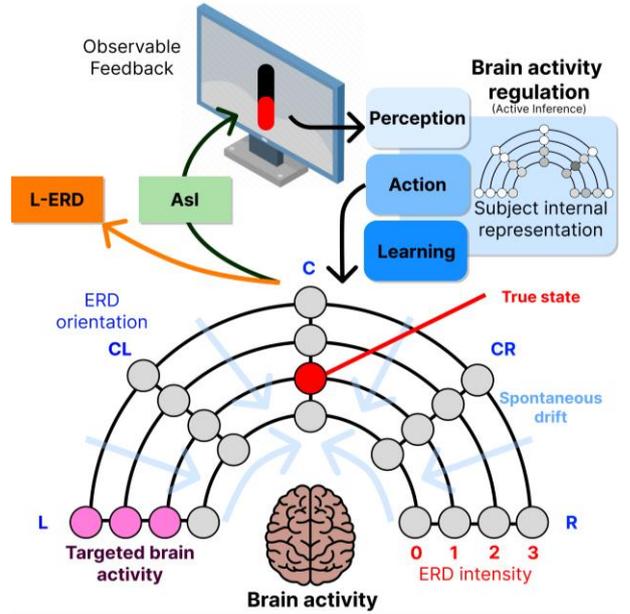

Figure 2: The modeled Motor Imagery intensity / orientation training. The agent internal representation drives the brain activity based on the feedback provided.

*Transitions:* At each timestep t, the subject's true states evolve depending on previous state value $\hat{s}_{t-1}$ and mental actions $u_{t-1}$. During the actual task, agents could potentially explore and use a large number of mental strategies (attentional / sensorial exercises, relaxation efforts, etc.), among which only a limited amount would prove "effective" and allow the subject to control their mental state with a probability $p_{effect} = 0.99$. Since mental actions are poorly understood, we assumed a synthetic topological state space that statisfies the three as following constraints:
- *Continuity:* for a specific state factor ($i$ / $\alpha$), the mental states could only move from value k to adjacent (or same) values {k-1, k , k+1}.
- *Invariability:* for a specific state factor, the effect of "effective" actions was independent from the occupied state.
- *Resting states:* to reflect the natural tendency of Motor imagery intensity to return to a resting state, non "effective" actions pulled the mental state of the subject towards this resting state with probability $p_{decay} = 0.1$. The resting states were $\hat{\imath} = 0$ for MI intensity and $\hat{\alpha} = C$ (center) for MI orientation.

For each state factor ($i$ / $\alpha$), $N_{up} = N_{down} = 1$ action were "effective" and allowed the subject to control their mental states. $N_{neutral} = 10$ actions were non "effective" and resulted in spontaneous drift towards the resting state.

*Subject priors:* the agents entertained representational priors about the BCI loop before starting the training. This included belief about the feedback modality (also called the 'likelihood model' **a**) and beliefs about the effect of their mental actions (**b**).

We assumed biased agents. We model them as having the expectation that high levels of motor imagery intensity would lead to higher feedback levels. This is in fact misleading but fits with the initial instructions they actually received in this experiment: "to perform Motor Imagery". This assumption is supported with further arguments in the discussion. The agents' model of the feedback was initiated using the Dirichlet conjugate prior for the categorical likelihood **a:**

$$\mathbf{a_0} = c_a \mathbf{1} + s_a Cat(N(i. AsI(\alpha); \sigma_{model})) \quad \text{Eq.4}$$

Where $c_a$ and $s_a$ are the initial concentration and confidence parameters, which we set to 1 and 100, respectively. This means that subjects were very confident that the feedback actually reflects (albeit with some noise) their mental imagery level. $AsI$ is the asymmetry index previously formulated and $\sigma_{model}$ is a noise term encoding subject's prior confidence in the feedback modality. It was set to 0.5.

Finally, subject prior beliefs about their mental actions were set as the combination of three terms: a prior concentration parameter $c_b$ indicating how much new evidence is needed for the subjects to change their prior beliefs, a 'stickiness' parameter $s_b$ that encodes subject's belief about actions not affecting their mental state, and an initial mental action confidence vector $b_{pre}$ that encodes previous knowledge about the effect of their mental actions. Importantly, $b_{pre}$ is a vector with one value for each state factor ($i$ / $\alpha$). The initial mental action model of the agents was thus, for each state factor:

$$\mathbf{b_0} = c_b \mathbf{1} + s_b \mathbf{Id} + b_{pre} \mathbf{B} \quad \text{Eq.5}$$

With **Id** the identity matrix. Simulations were conducted with $c_b = 1.0$ and $s_b = 1.0$ (i.e. subjects were opened to new evidence regarding their mental strategies). Of course, subjects started the training with relatively low values of $b_{pre}$, as high values of the parameter would render the training useless (this would mean the subject was already knowing how to perform the task optimally).

*Goals & simulations:* Using this simple model of self-regulation, our goal was to predict training outcome depending on the individual priors of each subject. Therefore, several families of agents were instantiated with various initial mental imagery familiarity levels. We demonstrate how these priors affect the way subjects learn how to perform the task and the evolution of the overall quality of their mental imagery models. To that end, we conducted simulations of agents performing Active Inference using the parametrized graph parameters **a0**, **b0**, **A**, **B**. The process parameters used in these simulations are $\sigma_{process}$, $b_{pre}(i)$ and $b_{pre}(\alpha)$.

All simulations in this paper were conducted using *active_pynference*, a freely available Python package for running sophisticated inference schemes. The code used in these simulations is freely available at: https://github.com/Erresthor/ActivPynference_Public/blob/main/paper_scripts/paper_grazBCI/simulations.ipynb .

RESULTS

*Agents already familiar with MI:* Figure 3 illustrates the outcome of 10 simulated agents performing 10 trials each, starting with informed action priors $b_{pre}(i) = 1$ and $b_{pre}(\alpha) = 1$. These subjects thus started the training with high mental imagery control skills, rendering the training unnecessary. The feedback provided was noisy, but informative ($\sigma_{process} = 1.5$). The average simulated mental states (true ERD intensity and orientation) are shown as well as the provided feedback (green). These can be compared to the corresponding performances of neurofeedback subjects from [16] shown below for a few subjects (Figure 3.A). The quite large mismatch between the empirical and simulated time series suggest that subjects entertained less precise action priors. Interestingly, the agents quickly learned to maintain a weak ERD strength while correctly lateralizing their ERDs, leading to less effortful, more optimal behavior.

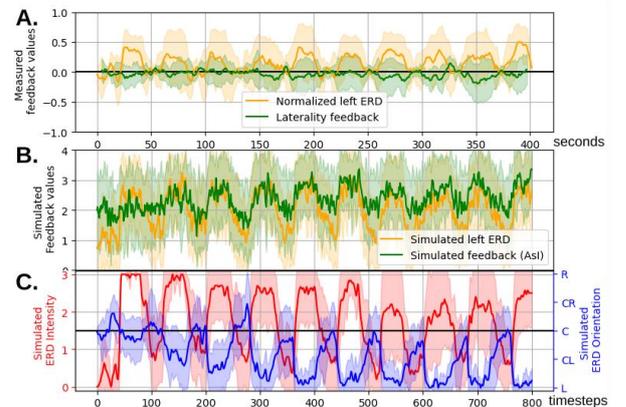

Figure 3: First empirical trials from [16] (A.) compared with the simulated laterality feedback and left ERD (B.) and motor imagery states (C.) from 10 simulated agents with high initial motor imagery control.

*Agents initially unable to perform MI lateralization:* Another class of agents was instantiated who were initially unable to produce lateralized motor imagery. They had no priors on how to control the orientation of their ERD ($b_{pre}(\alpha) = 0.0$), but had some poor priors on how to control their intensity ($b_{pre}(i) = 0.1$). They thus had to fully rely on the feedback to learn these transitions. To facilitate their training, a fairly reliable biomarker was assumed ($\sigma_{proc} = 0.5$). The training results are show in Figure 4. Overall, agents managed to reliably produce a lateralized ERD, although after quite a long training.

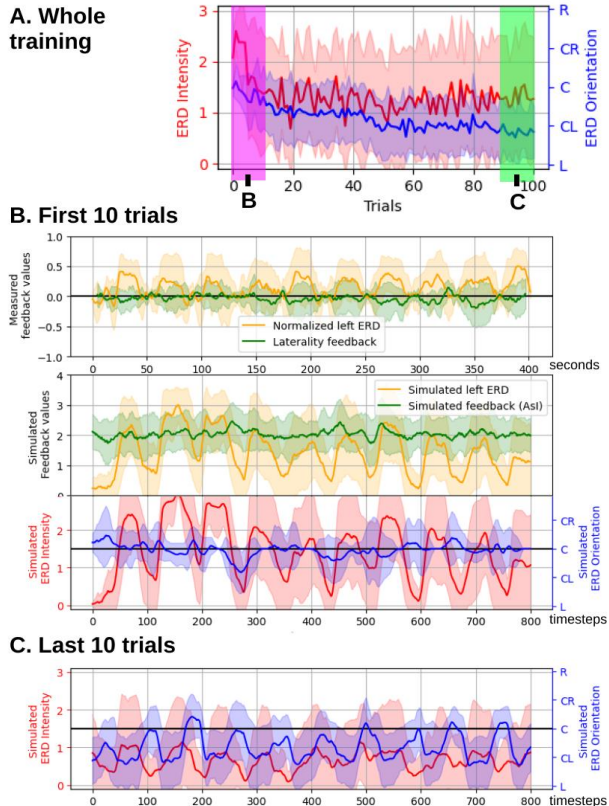

Figure 4: Agents with no prior knowledge of ERD lateralization performed 100 simulated neurofeedback trials. We show their average ERD strength (red) and orientation (blue) across the training (A) and at specific points of the training (B, C) The similarity between the simulated initial MI levels and the empirical observations (B) suggests that this set of parameters better matches the data than the over-optimistic previous simulations.

*Agents with mixed prior abilities:* Finally, 21 x 21 group of 10 agents with intermediate MI lateralization priors were simulated. Each group had a different pair of parameter values $\{b_{pre}(i), b_{pre}(\alpha)\}$, set between 0 and 2. This reflected individual differences in subjects starting BCI training with different Motor Imagery prior experience. The feedback provided was very noisy ($\sigma_{proc} = 1.5$). Figure 5 shows the evolution of average Motor Imagery performance in each group of subjects, at the start of training and at the end. Our simulations reveal counter-intuitive training effects, such as poor training results from subjects initially well versed in their ability to lateralize their ERDs but lacking the ability to reliably perform an ERD (e.g. subjects who misinterpret Motor Imagery by performing right hand visual instead of kinesthetic motor imagery). Conversely, subjects who were very good at performing mental imagery but lacked control over their MI laterality tended to benefit from training and managed to learn how to direct their attention, despite the noisy feedback.

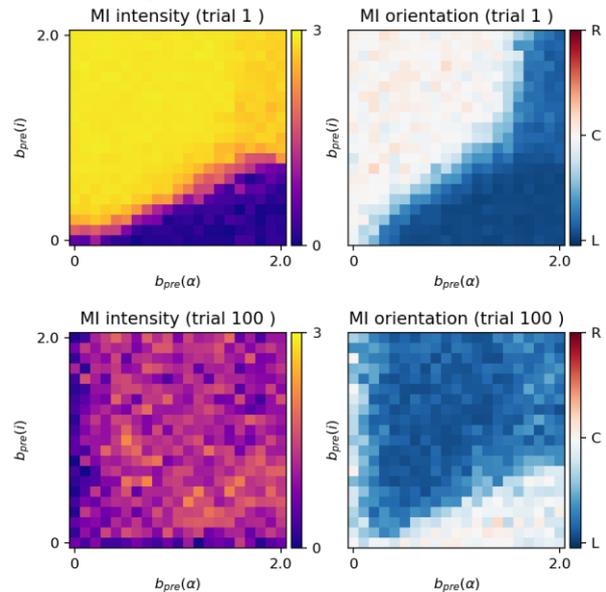

Figure 5: Simulated motor imagery performance before (top) and after (bottom) neurofeedback depending on initial experience $\boldsymbol{b_{pre}(i)}$ (y-axis) and $\boldsymbol{b_{pre}(\alpha)}$ (x-axis).

DISCUSSION

The reported simulations provided an account of Motor Imagery training using Neurofeedback for various groups of subjects parametrized mostly by their past experience with Motor Imagery: (i) subjects familiar with motor imagery and who had good initial priors, (ii) subjects with poor initial ability to lateralize their ERD and had to learn from scratch, and (iii) intermediate subjects who started with mixed priors about MI laterality and strength, but had to finetune them in order to perform the task efficiently.

Simulations showcased very different training curves and general subject classes that would more or less benefit from the training depending on their initial situation. They illustrated the crucial role of subject's prior skills (i.e. previous experience), expectations about the feedback, training and beliefs following task instructions. Subjects starting training with uninformed priors performed poorly. This was in part due to the sparse feedback modality (low temporal resolution / low dimensionality) which made learning from scratch a very tricky task. This suggests that reducing the amount of targeted mental dimensions may be instrumental to guarantee successful training [6]. The lackluster ability of the subjects when they had to build a model of interaction from scratch also suggests that more basic learning mechanisms such as classical (model free) Reinforcement Learning may play a significant role in the initial phase of the training, with a more complex representational learning taking over later on [13].

The proposed framework is very general and flexible enough to capture a large variety of experimental paradigms. For instance, the multidimensional feedback based learning implemented in [16] may be modelled by agents learning simultaneously several sensory mappings of the same internal dynamical state.

CONCLUSION

This paper presents a computational account of neurofeedback/BCI Motor Imagery training using the Active Inference framework. Preliminary simulations reveal that the Active Inference framework has great potential to provide an account of individual self-regulation dynamics. Future work will consist in fitting alternative instantiations of such models to actual data in order to demonstrate the validity of this approach to disentangle between learning profiles and identify individual traits for BCI learning curves and empirically observed neurophysiological dynamics.